\documentclass[%
 aip,
 amsmath,amssymb,
 reprint,%
]{revtex4-1}

\usepackage{graphicx}
\usepackage{dcolumn}
\usepackage{bm}

\usepackage[utf8]{inputenc}
\usepackage[T1]{fontenc}
\usepackage{mathptmx}
\usepackage{etoolbox}

\usepackage[unicode=true,colorlinks=true,citecolor=blue]{hyperref}

\makeatletter
\def\@email#1#2{%
 \endgroup
 \patchcmd{\titleblock@produce}
  {\frontmatter@RRAPformat}
  {\frontmatter@RRAPformat{\produce@RRAP{*#1\href{mailto:#2}{#2}}}\frontmatter@RRAPformat}
  {}{}
}%
\makeatother

\newcommand{\divv}{\mathop{\rm div}\nolimits}
\newcommand{\grad}{\mathop{\rm grad}\nolimits}
\newcommand{\Tr}{\mathop{\rm Tr}\nolimits}
\makeatletter
\newsavebox{\@brx}
\newcommand{\llangle}[1][]{\savebox{\@brx}{\(\m@th{#1\langle}\)}%
  \mathopen{\copy\@brx\kern-0.5\wd\@brx\usebox{\@brx}}}
\newcommand{\rrangle}[1][]{\savebox{\@brx}{\(\m@th{#1\rangle}\)}%
  \mathclose{\copy\@brx\kern-0.5\wd\@brx\usebox{\@brx}}}
\makeatother

\usepackage[normalem]{ulem}
\usepackage[dvipsnames]{xcolor}

\newcommand{\addSivan}[1]{\textcolor{blue}{#1}}

\newcommand{\addZakhar}[1]{\textcolor{Emerald}{#1}}

\begin{document}


\title[Phonon-induced exciton weak localization]{Phonon-induced exciton weak localization\\ in two-dimensional semiconductors}
\author{M.M. Glazov}
 \email{glazov@coherent.ioffe.ru}
\affiliation{ 
Ioffe Institute, 194021, St. Petersburg, Russia
}%

\author{Z.A. Iakovlev}
\affiliation{ 
Ioffe Institute, 194021, St. Petersburg, Russia
}%

\author{S. Refaely-Abramson}
\affiliation{Department of Molecular Chemistry and Materials Science, Weizmann Institute of Science, Rehovot 7610001, Israel}

\date{\today}

\begin{abstract}
We study theoretically the contribution of quantum effects to the exciton diffusion coefficient in atomically thin crystals. It is related to the weak localization caused by the interference of excitonic wavefunctions on the trajectories with closed loops. Due to a weak inelasticity of the exciton-phonon interaction the effect is present even if the excitons are scattered by long-wavelength acoustic phonons. We consider exciton interaction with longitudinal acoustic phonons with linear dispersion and with flexural phonons with quadratic dispersion. We identify the regimes where the weak localization effect can be particularly pronounced. We also briefly address the role of free charge carriers in the exciton quantum transport and, within the self-consistent theory of localization, the weak localization effects beyond the lowest order.
\end{abstract}

\maketitle

\section{\label{sec:intro}Introduction}

Electron-phonon interaction is a key to understand kinetic and transport properties of charge carriers and excitons in semiconductors.\cite{gantmakher87,toyozawa_2003} It governs momentum and energy relaxation processes, intervalley transitions in multivalley systems and, via the spin-orbit coupling, spin relaxation and decoherence.\cite{RevModPhys.89.015003} In the strong coupling regime, polarons -- bound quasiparticles of electronic excitations and lattice vibrations -- can be formed\cite{Alexandrov2010,emin2013polarons} strongly influencing exciton relaxation mechanisms and lifetimes.

In emergent two-dimensional (2D) semiconductors like graphene~\cite{Geim2007} and transition-metal dichalcogenide monolayers (TMDC MLs)~\cite{Mak:2010bh,Splendiani:2010a} the strength of the electron-phonon interaction is increased providing significant impact on the optical and transport properties of these nanosystems.\cite{PhysRevLett.119.187402,shree2018exciton} The Raman scattering of light in atomically thin crystals mediated by the phonons provides important information about the electron bandstructure,\cite{ferrari06} mechanisms of electron-phonon~\cite{ferrari07_2} and electron-electron interactions.\cite{PhysRevB.78.125418,Eliseyev:2019ws} Enhanced electron-phonon interaction allows one to study phonon cascades in monolayers via the resonant Raman processes.\cite{2020arXiv200705369P} Phonons also affect  spin-valley dynamics of the charge carriers and excitons.\cite{Kioseoglou,Song:2013uq,PhysRevLett.115.117401,Miller:2019aa} Another important property of phonons in 2D semiconductors is the presence of soft out-of-plane polarized acoustic modes,\cite{lifshits1952low} that provide\addZakhar{s} anomalous contributions to thermodynamic and kinetic properties of such systems~\cite{nelson:2004aa,Katsnelson2020} and give\addZakhar{s} rise to unconventional polaron effects.\cite{PhysRevB.82.205433,https://doi.org/10.1002/andp.202000339,2022arXiv220212143I}

Importantly, exciton-phonon coupling controls transport properties of the quasi-particles in high-quality hBN encapsulated TMDC MLs: It is the dominant relaxation and dephasing channel over a wide range of temperatures.\cite{Selig:2016aa} The key parameter here is the exciton diffusion coefficient $D$ which can be measured optically.\cite{Unuchek:2018aa,Cadiz:2018aa,PhysRevLett.120.207401,zipfel2019exciton,Wang:2019aa,Uddin:2020wq,Akmaev:2021up} Within the semiclassical approach the exciton diffusion coefficient is determined by the product of the squared thermal velocity and the scattering time. It makes $D$ temperature-independent at sufficiently low temperatures where the scattering is dominated by the long wavelength acoustic phonons with linear dispersion.\cite{PhysRevLett.124.166802} Thus excitons in 2D semiconductors are particularly promising as a platform to study non-classical effects beyond this picture, particularly, weak localization,\cite{Gorkov:WL} via the temperature dependence of the diffusion coefficient. The weak localization effect results from the exciton interference at the trajectories containing closed loops\cite{1977ZhETF..72.2230I,golubentsev,PhysRevA.36.5729,PhysRevB.36.5663,PhysRevLett.58.2106,1988JETPL..47..259D,arseev98} and gives rise to a temperature-dependent non-classical contribution to their diffusion coefficient.\cite{PhysRevLett.124.166802} The signatures of non-classical diffusion of excitons have been recently observed in WSe$_2$ monolayer.\cite{PhysRevLett.127.076801}

In this paper, we study the temperature dependence of the weak localization induced correction to the exciton diffusion coefficient in monolayer semiconductors in the regime where exciton scattering by long-wavelength  phonons is dominant. We address two important situations where (i) excitons interact with longitudinal acoustic phonons with linear dispersion and (ii) excitons are scattered by the flexural -- out-of-plane -- phonons with quadratic dispersion. In the second situation we consider both the two-layer structures with relatively strong interaction of excitons with flexural phonons\cite{https://doi.org/10.1002/andp.202000339} and the single-layer systems where the single-phonon processes are forbidden and exciton-two-phonon interactions are the most important.\cite{Katsnelson2020} We also discuss effects of the phonon damping on the exciton weak localization. Finally, motivated by the interest in the physics and transport properties of Bose-Fermi mixtures,\cite{PhysRevLett.104.106402,PhysRevX.9.041019,PhysRevLett.123.095301,wagner:trions} we  analyze the effect of free charge carriers on the exciton weak localization and its interplay with exciton-phonon interactions. In the end of the paper, we briefly discuss the quantum contributions to the exciton diffusion coefficient beyond the first-order weak localization corrections.

\section{Model}

\subsection{General approach to exciton transport}

We are interested in the model perspective of the exciton transport in two-dimensional semiconductors. We start with a general model providing a link between an \emph{ab initio} description of excitonic bandstructure and simplified picture of the diffusion of quasiparticles. Let us consider an ideal periodic crystal at a zero temperature and present the exciton Bloch functions in the following form:
\begin{multline}
\label{Psi}
\Psi_{\bm k,N}(\bm r_e, \bm r_h) = \frac{e^{\mathrm i \bm k \bm r}}{\sqrt{\mathcal S}}\Phi_{\bm k,N}(\bm r_e, \bm r_h) \\
= \sum_{\bm q, c,v}A^{N, \bm k}_{\bm q;c,v}\psi_{\bm q+ \bm k,c}(\bm r_e)\psi_{\bm q,v}(\bm r_h).
\end{multline}
Here $\bm k$ is the exciton translational motion wavevector, $\bm r_e$, $\bm r_h$ are the electron and hole position vectors, $\bm r$ is the exciton center of mass coordinate, $\mathcal S$ is the normalization area, $\mathcal S=1$ in what follows. Equation~\eqref{Psi} follows from the translational invariance of the crystal, correspondingly, the excitonic Bloch amplitude $\Phi_{\bm k,N}(\bm r_e, \bm r_h)$ is recast as a linear combination of $\psi_{\bm q,c}(\bm r_e)$, $\psi_{\bm q,v}(\bm r_h)$, the electron and hole Bloch functions, with decomposition coefficients $A^{N, \bm k}_{\bm q;c,v}$ that can be found from the Bethe-Salpeter equation,\cite{Rohlfing1998, Qiu2015, Cudazzo2016} see also Refs.~\onlinecite{birpikus_eng,PhysRevB.102.155305}. In Eq.~\eqref{Psi} we use electron-electron representation and the summation is carried out over all unoccupied bands $c$ and occupied bands $v$. The multiindex $N$ encorporates all relevant quantum numbers, including the bands from which exciton originates and quantum numbers describing relative motion of the electron and hole in the exciton.


In the absence of exciton-phonon interaction and defects the system is translationally invariant and  $\Psi_{\bm k,N}(\bm r_e, \bm r_h)$ are the eigenfunctions of the system (more precisely, functions $\Psi_{\bm k,N}(\bm r_e, \bm r_h)$  are the eigenfunctions in the sector of two-particle, electron-hole, excitations of the two-dimensional semiconductor) and their temporal evolution is trivial and described by the time-dependent factors $\exp{(-\mathrm i \varepsilon_{\bm k,N} t/\hbar)}$, where $\varepsilon_{\bm k,N}$ is the exciton dispersion. To provide an analogy with a classical picture of a particle propagating in a media, it is instructive to introduce the wavepackets formed from the these basic functions~\eqref{Psi} as
\begin{equation}
\label{packet}
\Psi = \sum_{\bm k} e^{\mathrm i \bm k \bm r} C_{\bm k}(\bm k_0;N) \Phi_{\bm k,N}(\bm r_e, \bm r_h),
\end{equation}
with the decomposition coefficients $C_{\bm k}(\bm k_0;N)$ which are non-zero for $\bm k$ in the vicinity of a given ``central'' wavevector $\bm k_0$. The wavepackets are localized in the real and momentum spaces. The wavepackets~\eqref{packet} propagate with the group velocity 
\begin{equation}
\label{group:v}
\bm v_{\bm k_0} = \left.\frac{1}{\hbar} \frac{\partial \varepsilon_{\bm k,N}}{\partial \bm k}\right|_{\bm k = \bm k_0},
\end{equation}
and weakly spread in the real space as the wavefunctions in Eq.~\eqref{packet} are not the exact eigenstates of the system.

\begin{figure}[b]
    \centering
    \includegraphics[width=0.8\linewidth]{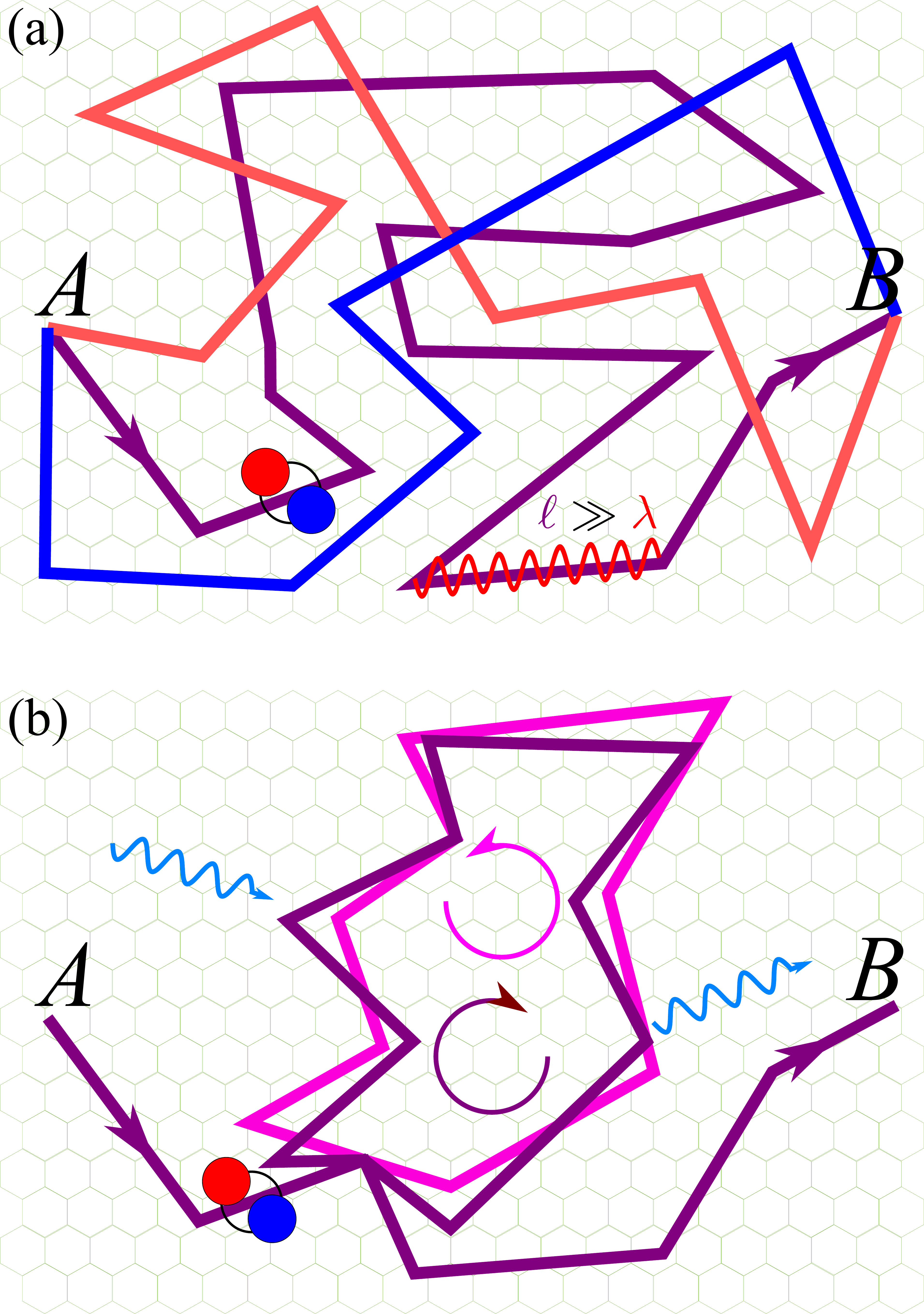}
    \caption{Schematics of diffusive propagation of excitons in 2D semiconductor. (a) Quasi-classical trajectories from $A$ to $B$ where the quantum interference under condition~\eqref{applicability} is unimportant. (b) Trajectory with a closed loop where interference plays a role. Clock- and counterclock-wise loops are slightly shifted due to the phonon propagation that result in exciton dephasing.}
    \label{fig:illustr}
\end{figure}

In the presence of static defects or phonons the wavefunctions~\eqref{Psi} are generally no longer eigenstates of the system: The translational invariance is broken, consequently, the center of mass momentum $\bm k$ is not a good quantum number anymore. In the case of a static disorder one can, at least in principle, find new exact eigenfunctions of excitons in the form $\sum_{\bm q, \bm q', c,v}A_{\bm q,\bm q';c,v}\psi_{\bm q',c}(\bm r_e)\psi_{\bm q,v}(\bm r_h)$ with certain coefficients $A_{\bm q,\bm q';c,v}$ and study their dynamics. By contrast, in the case of exciton-phonon interaction it is not possible since the proper description should also explicitly include phonon degrees of freedom. However, if exciton-phonon and exciton-defect interaction is not too strong, namely, if in the relevant range of energies defect-induced exciton localization does not influence their diffusivity and the formation of polarons is inefficient, the wavefunctions~\eqref{Psi} serve as a convenient basis set for the problem of exciton-phonon or exciton-defect scattering. Correspondingly, the wavepacket description of exciton dynamics in terms of functions~\eqref{packet} approximately holds at the time scale $t\lesssim \tau$, where $\tau$ is the characteristic scattering time: During the time $\tau$ exciton wavevector $\bm k$ significantly changes; it will be specified later in Sec.~\ref{sec:semiquant}. Correspondingly, in the first approximation, the exciton wavepacket propagates following semiclassical trajectories depicted in Fig.~\ref{fig:illustr}(a). Such picture is valid provided that the exciton mean free path $\ell \sim v_{\bm k_0} \tau$ is much longer than the exciton de Broglie wavelength $\lambda \sim k_0^{-1}$:
\begin{equation}
\label{applicability}
\ell \gg \lambda \quad\mbox{or} \quad \frac{k_B T \tau}{\hbar} \gg 1.
\end{equation}
Here we made use of the fact that for thermalized excitons their kinetic energy is given by the temperature. Additionally, it is assumed that both $\ell$ and $\lambda$ exceed by far the exciton Bohr radius.

Let us quantify now the exciton transport. In accordance with the state-of-the-art experimental works\cite{Unuchek:2018aa,Cadiz:2018aa,PhysRevLett.120.207401,zipfel2019exciton,Wang:2019aa,Uddin:2020wq,Akmaev:2021up} we study exciton propagation on the length scales exceeding by far both $\lambda$ and $\ell$ and on the time scale $t \gg \tau$. In that case, the exciton dynamics is controlled by the continuity equation
\begin{subequations}
\label{propagation:eqs:gen}
\begin{equation}
 \label{cont}
 \frac{\partial n(\bm r,t)}{\partial t} + \divv{\bm i(\bm r,t)} + \frac{n}{\tau_r} =0,
\end{equation}
and the material relation 
\begin{equation}
    \label{mater}
    \bm i(\bm r,t) = - D \grad n(\bm r,t).
\end{equation}
Here $n(\bm r,t)$ is the total density of excitons, $\bm i$ is the exciton flux (exciton current density), $\tau_r$ is the exciton lifetime (it includes contributions from all exciton decay processes, including radiative and non-radiative recombination, dissociation, etc.), and $D$ is the exciton diffusion coefficient. Combining Eqs.~\eqref{cont} and \eqref{mater} we arrive at the diffusion equation
\begin{equation}
\label{diffusion}
 \frac{\partial n(\bm r,t)}{\partial t}  + \frac{n}{\tau_r} = D \Delta n(\bm r,t),
\end{equation}
\end{subequations}
where $\Delta$ is 2D-Laplacian. 
Equations~\eqref{propagation:eqs:gen} are valid on time and length scales that are much longer than microscopic scales of the problem, these equations are general and hold for any regime of exciton transport. The description based in Eqs.~\eqref{propagation:eqs:gen} is not limited to a semiclassical propagation picture and can be applied, e.g., if excitons were localized and hop between the localization sites emitting and absorbing phonons. Thus, the key parameter describing exciton transport is $D$, the exciton diffusion coefficient.

\begin{figure}[b]
    \centering
    \includegraphics[width=0.99\linewidth]{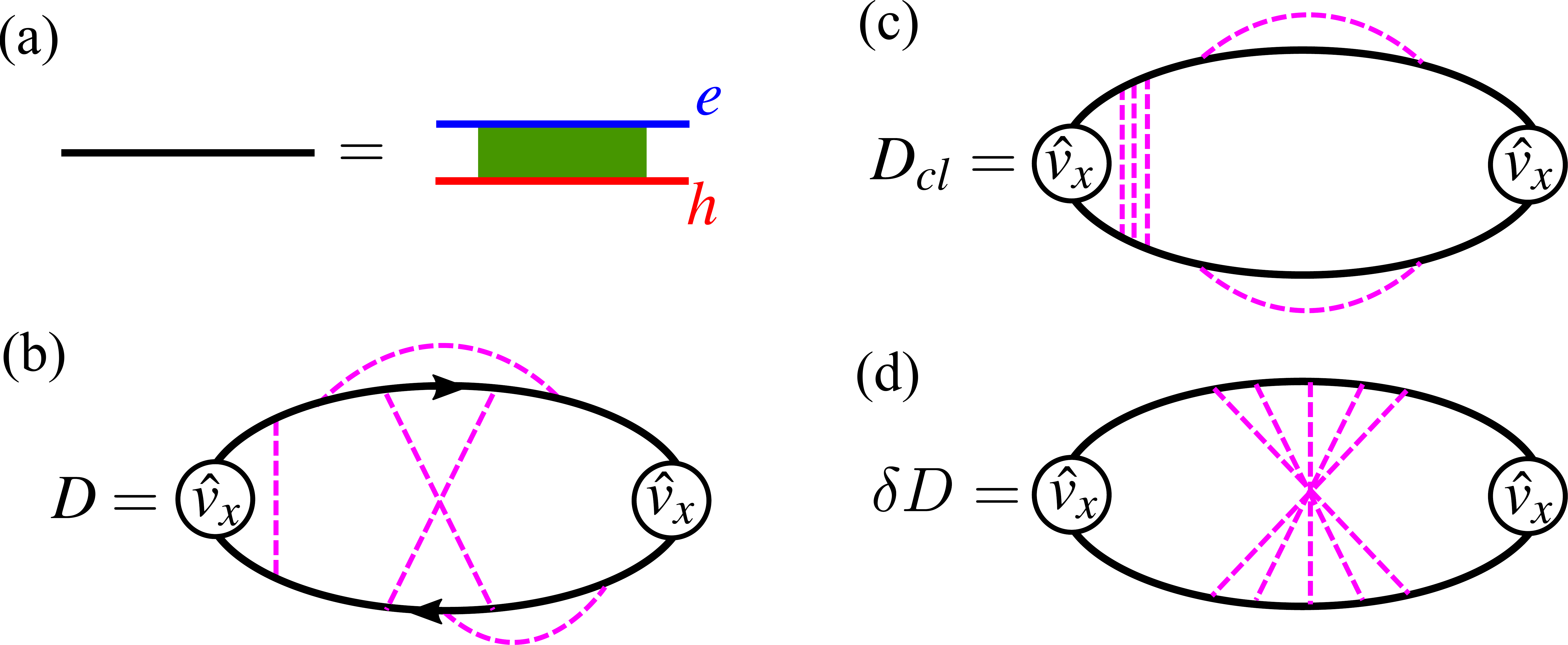}
    \caption{Diagrammatic approach to the exciton diffusion coefficient. (a) Exciton Greens function: the two-particle electron (e) and hole (h) Greens function with account for all possible Coulomb interactions (green box). (b) Generic diagram describing the diffusion coefficient with $\hat v_x$ being the velocity vertex, black arrows being the exciton Greens functions and magenta dashed lines show the correlator of the random potential induced by the static defects or phonons. (c) Semiclassical diffusion coefficient, Fig.~\ref{fig:illustr}(a). (d) Maximally crossed diagram describing contribution due to the weak localization effect, Fig.~\ref{fig:illustr}(b). }
    \label{fig:d:diag}
\end{figure}

Making use of the fluctuation-dissipation theorem the diffusion coefficient $D$ can be presented in the form of the velocity autocorrelation function
\begin{equation}
\label{D:gen:vxvx}
D = \int_0^{\infty} \llangle \hat v_x(t) \hat v_x(0) \rrangle\, dt,
\end{equation}
where $\hat v_x(t)$ is the $x$-component velocity operator in the Heisenberg representation and the double angular brackets $\llangle \ldots \rrangle$ denote averaging over the equilibrium distribution of excitons and scattering. It is convenient to calculate the autocorrelation function~\eqref{D:gen:vxvx} using the Greens function method and diagram technique. In order to introduce the exciton Greens function it is instructive, instead of using the basic Bloch amplitudes $\Phi_{\bm k, N}$ in Eq.~\eqref{Psi} that depend explicity of the exciton wavevector $\bm k$, to decompose the exciton Bloch functions $\Psi_{\bm k,N}$  over the Bloch amplitudes with the fixed $\bm k = \bm 0$ as follows~\cite{birpikus_eng}
\begin{equation}
\label{Psi:1}
\Psi_{\bm k,N}(\bm r_e, \bm r_h)= e^{\mathrm i \bm k \bm r} \sum_{N'} F_{N'}^N(\bm k) \Phi_{\bm 0, N'}(\bm r_e, \bm r_h),
\end{equation}
where $F_{N'}^N(\bm k)$ are the decomposition coefficients. We introduce the Hamiltonian $\hat{\mathcal H}(\bm k)$, as the matrix in the space of exciton quantum number $N, N'$, that governs the exciton bands such that
\begin{equation}
\label{Ham}
\sum_{N'} \mathcal H_{NN'}(\bm k) F_{N'}^N(\bm k) = F_{N}^N(\bm k) \varepsilon_{\bm k, N}.
\end{equation}
The Hamiltonian $\hat{\mathcal H}$ allows us to express the bare retarded and advanced excitons Greens functions as
\begin{equation}
\label{G:R:A}
\hat{\mathcal G}^{R/A}_{\bm k}(\varepsilon) = \left[\varepsilon - \mathcal H(\bm k) \pm \mathrm i 0\right]^{-1}, 
\end{equation}
and present the diffusion coefficient in Eq.~\eqref{D:gen:vxvx} in the form\cite{PhysRevLett.124.166802}
\begin{equation}
    \label{D:gen:GRGA}
    D = \frac{\hbar\sum_{\bm k} \int_{-\infty}^\infty \frac{d\varepsilon}{2\pi} \Tr\{\hat n_{\bm k} \langle \hat{v}_x\hat{\mathcal G}^{R}_{\bm k}(\varepsilon) \hat{v}_x\hat{\mathcal G}^{A}_{\bm k}(\varepsilon) \rangle \}}{\sum_{\bm k} \Tr\{\hat n_{\bm k}\}}.
\end{equation}
Here $\hat n_{\bm k}$ is the equilibrium density matrix of excitons, the trace is taken over $N$ and angular brackets denote the averaging of the Greens function product over all possible scattering acts. In the diagram representation, Fig.~\ref{fig:d:diag}, the diffusion coefficient is expressed by the loop of two excitonic Greens functions [solid black lines, Fig.~\ref{fig:d:diag}(a)] dressed by the disorder correlation functions [dashed magenta lines in Fig.~\ref{fig:d:diag}(b-d)]. In the case of exciton-phonon scattering relevant for the following dashed lines represent the phonon propagators multiplied by the products of the exciton-phonon interaction matrix elements. We stress that despite the fact that we used the basis of the states with a given exciton wavevector $\bm k$ to construct the Greens functions~\eqref{G:R:A} and express the diffusion coefficient in Eq.~\eqref{D:gen:GRGA}, Eqs.~\eqref{D:gen:vxvx} and \eqref{D:gen:GRGA} can be written in any appropriate basis and cover all possible regimes of exciton propagation, from semiclassical one where excitons are almost free and rarely scatter, to the hopping regime where the excitons are strongly localized at the disorder.

\subsection{Semiclassical and quantum effects in a single band model}\label{sec:semiquant}



Formally, Eq.~\eqref{D:gen:GRGA} describes exciton diffusion for arbitrary bandstructure and any strength of the exciton-phonon scattering, the calculation of all diagrams depicted in Fig.~\ref{fig:d:diag}(b-d) is extremely complicated. Thus, several approximations and simplifications are implemented and we consider them in more detail below. The first issue with evaluation of exciton diffusion via Eq.~\eqref{D:gen:GRGA} is the presence of multiple excitonic states enumerated by the index $N$: generally, all scattering processes $\bm k, N \to \bm k',N'$ with arbitrary initial and final states are possible. However, at not too high temperatures only one or several close-lying exciton bands are involved, e.g., for dark excitons in WSe$_2$ monolayers\cite{PhysRevLett.127.076801} or interlayer excitons at WSe$_2$/MoSe$_2$ heterostructures.\cite{Barr2022,PhysRevLett.125.255301}  
It allows us for the sake of simplicity to use a single band approximation fixing $N$ hereafter. Accordingly, we omit $N$-dependence of all relevant quantities.

The second issue is related with a variety of the diagrams. However, in the semiclassical regime of exciton propagation, where the exciton mean free path $\ell$ exceeds by far its de Broglie wavelength $\lambda$,  Eq.~\eqref{applicability}, the various types of the diagrams can be readily classified and the most relevant diagrams have a transparent physical interpretation. The ratio $\lambda/\ell \ll 1$ is a small parameter of our theory. Indeed, at $\ell \gg \lambda$, the quantum-mechanical interference of different trajectories connecting the initial point $A$ and the final point $B$ can be disregarded, and the diffusion coefficient takes the simple form
\begin{equation}
    \label{diffusion:class}
    D_{cl} = \int_0^\infty \exp{\left( -\frac{\varepsilon}{k_B T}\right)} D(\varepsilon) \frac{d\varepsilon}{k_B T},
\end{equation}
where the exponential factor accounts for the Boltzmann statistics of excitons with $T$ being the temperature and $k_B$ the Boltzmann constant, and $D(\varepsilon)$ is the energy-dependent diffusion coefficient expressed as
\begin{equation}
    \label{D:eps}
    D(\varepsilon) = \frac{\varepsilon}{M} \tau(\varepsilon).
\end{equation}
Here $M$ is the exciton translational motion mass and $\tau(\varepsilon)$ is the momentum relaxation time. For a quasi-elastic scattering regime relevant for the exciton interaction with acoustic phonons where exciton kinetic energy exceeds the phonon energy the momentum relaxation rate reads
\begin{equation}
    \label{transport:time}
    \frac{1}{\tau(\varepsilon_k)} = \sum_{\bm k'} W_{\bm k', \bm k} (1-\cos{\vartheta_{\bm k',\bm k}}) \delta(\varepsilon_k - \varepsilon_{k'}).
\end{equation}
In Eq.~\eqref{transport:time} $W_{\bm k', \bm k}$ is the rate of phonon-assisted transitions between the states $\bm k$ and $\bm k'$, $\vartheta_{\bm k',\bm k}$ is the scattering angle [$\cos\vartheta_{\bm k',\bm k} = \bm k\cdot \bm k'/(kk')$], $\varepsilon_k = \hbar^2 k^2/2M$ is the exciton dispersion, and the factor $1-\cos{\vartheta_{\bm k',\bm k}}$ ensures that the diffusion is controlled by the scattering processes where the momentum of the exciton significantly changes. Notably, the exciton dispersion can be non-parabolic.\cite{Qiu2015} We exclude this case here, but in general the exciton (momentum-dependent) effective mass can be evaluated for these scenarios as well,\cite{Qiu:2021to} as long as an analytical bands shape is maintained. We also introduce the quantum relaxation time (out-scattering time) $\tau_q(\varepsilon)$ as 
\begin{equation}
    \label{quantum:time}
    \frac{1}{\tau_q(\varepsilon_k)} = \sum_{\bm k'} W_{\bm k', \bm k}  \delta(\varepsilon_k - \varepsilon_{k'}).
\end{equation}
It differs from the transport or momentum relaxation time~\eqref{transport:time} by the absence of the transport factor $1-\cos{\vartheta_{\bm k',\bm k}}$, i.e., the rate $1/\tau_q$ accounts for all possible scattering processes, including forward scattering ones where the exciton momentum does not significantly change. 
Note that the time $\tau_q$ determines the ``optical'' coherence time of excitons, $\hbar/\tau_q(0)$ gives an estimate for the exciton linewidth in optical spectra.

Equations~\eqref{diffusion:class} -- \eqref{transport:time} can be readily derived from the kinetic equation~\cite{Perea-Causin:2019aa,PhysRevB.100.045426} for the exciton distribution function $n_{\bm k}$. Within the diagram approach it corresponds to the contribution of the diagrams where phonon lines do not cross, Fig.~\ref{fig:d:diag}(c). To that end, it is convenient to introduce the correlation function $\mathcal K(\bm r, t)$ of the potential $U(\bm r, t)$ induced by the phonons and its Fourier transform $\mathcal K_{\bm q,\omega}$ as\cite{PhysRevLett.124.166802}
\begin{multline}
\label{phonon:correlator:gen}
\mathcal K(\bm r_1 - \bm r_2, t_1 - t_2) = \langle U(\bm r_1, t_1) U(\bm r_2, t_2)\rangle, \\
    \mathcal K_{\bm q,\omega} = \int \mathcal K(\bm r, t) e^{-\mathrm i \bm q \bm r +\mathrm i\omega t}\, d\bm r \, dt.
\end{multline}
Hereafter we consider the relatively high-temperature regime where the phonon occupancy is large and the phonons can be treated semiclassically in a sense that phonon absorption and emission rates are the same. The correlator $\mathcal K_{\bm q,\omega}$ allows us to express the relaxation rates in the form
\begin{subequations}
\label{times:correlator:gen}
\begin{equation}
\label{momentum}
\frac{1}{\tau(\varepsilon)} = \frac{1}{\hbar^2} \sum_{\bm k'} (1-\cos{\vartheta_{\bm k',\bm k}})\mathcal K_{\bm k -\bm k', (\varepsilon-\varepsilon_{k'})/\hbar},
\end{equation}
for the momentum relaxation and
\begin{equation}
\label{quantum}
\frac{1}{\tau_q(\varepsilon)} = \frac{1}{\hbar^2} \sum_{\bm k'} \mathcal K_{\bm k -\bm k', (\varepsilon-\varepsilon_{k'})/\hbar},
\end{equation}
for the quantum relaxation rates, respectively. It is noteworthy that for quasi-elastic scattering processes studied here the transition rates can be also recast as
\begin{multline}
    \label{rates:K:alt}
    \left\{
    \begin{matrix}
    \tau^{-1}(\varepsilon)\\
    \tau_{q}^{-1}(\varepsilon)
    \end{matrix}
    \right\} 
    \\
    = \frac{M}{2\pi \hbar^3}\int_{-\infty}^\infty d\omega \left\langle  \left\{
    \begin{matrix}
    (1-\cos{\vartheta_{\bm k',\bm k}})\\
    1
    \end{matrix}
    \right\} \mathcal K_{\bm k -\bm k', \omega}\right \rangle.
\end{multline}
\end{subequations} 
Here the angular brackets denote averaging over the scattering angle $\vartheta_{\bm k',\bm k}$ at fixed absolute values $k=k' = \sqrt{2M\varepsilon/\hbar^2}$. Using the standard rules of the diagram technique\footnote{Most conveniently the calculation is performed in the Keldysh technique assuming the exciton occupancies are small} we obtain the Greens functions of excitons interacting with phonons
\begin{equation}
    \label{G:R:A:ph}
    \langle \hat{\mathcal G}^{R/A}_{\bm k}(\varepsilon)\rangle = \frac{1}{\varepsilon - \varepsilon_{\bm k} \pm \mathrm i \frac{\hbar}{2\tau_q(\varepsilon)} \pm \mathrm i \frac{\hbar}{2\tau_r}}.
\end{equation}
These expressions correspond to account for the diagrams in Fig.~\ref{fig:d:diag}(c) where the initial and final points of the dashed line start and end on the same Greens function (top or bottom). In Eq.~\eqref{G:R:A:ph} we disregard the renormalization of the exciton dispersion $\varepsilon_k$ due to the polaronic effects (i.e., due to the real part of the corresponding self-energies) as well as the contributions of diagrams with the crossed phonon lines, the latter are small under condition~\eqref{applicability}. Hereafter we assume that $\tau_r \gg \tau_q, \tau$. Further, using Eq.~\eqref{D:gen:GRGA} we can readily calculate the diffusion coefficient and arrive at Eqs.~\eqref{diffusion:class} and \eqref{D:eps}. The allowance for the vertical lines connecting the Greens functions in Fig.~\ref{fig:d:diag}(c) results in the vertex correction and the factor $1-\cos{\vartheta_{\bm k', \bm k}}$ in the definition of the time $\tau(\varepsilon)$, Eqs.~\eqref{transport:time} or \eqref{momentum}, entering the diffusion coefficient \eqref{D:eps}.

\begin{figure}[h]
    \centering
    \includegraphics[width=0.5\linewidth]{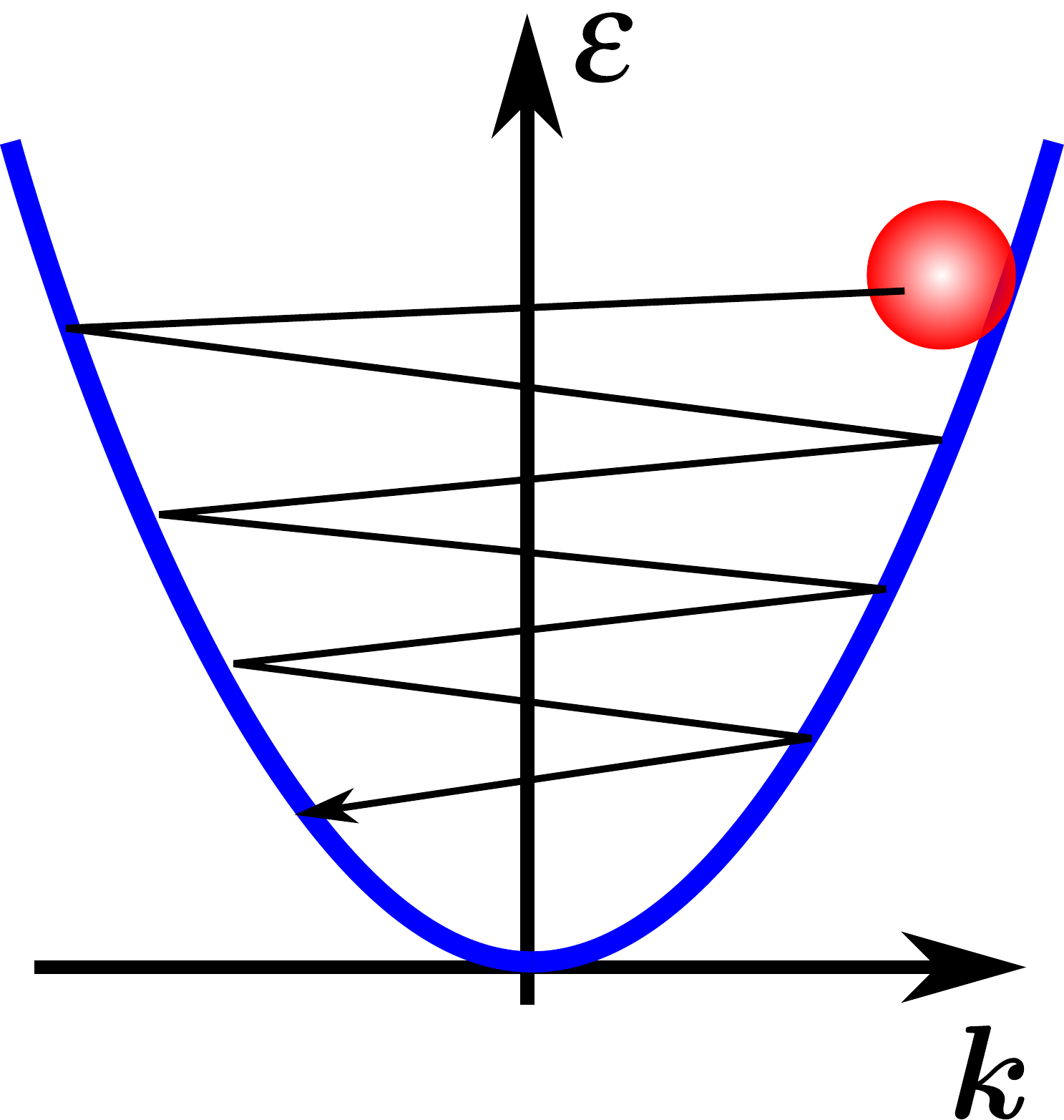}
    \caption{Sketch of the momentum and energy relaxation at the quasi-elastic scattering. The phonon momentum is of the same order and the exciton momentum, while the phonon energy is much smaller than the exciton energy.}
    \label{fig:times}
\end{figure}

We recall that the exciton interaction with acoustic phonons is quasi-elastic: The energy transferred in the act of exciton-phonon scattering is small compared with the mean exciton energy:  $\hbar\Omega_{\bm k'- \bm k} \ll \varepsilon_k,\varepsilon_k'$, where $\Omega_{\bm q}$ is the phonon dispersion. As a result, the exciton looses its energy in small portions, i.e., diffusively, as shown in Fig.~\ref{fig:times}. The mean squared variation of exciton energy in the scattering act reads
\begin{equation}
    \label{dOmega2}
    \hbar^2\delta\omega^2 = \frac{\tau_q(\varepsilon) M}{2\pi \hbar^3} \int_{-\infty}^\infty d\omega \langle (\hbar\omega)^2\mathcal K_{\bm k' - \bm k,\omega}\rangle.
\end{equation} 
This quantity will be useful in what follows.

In the semiclassical approach the exciton propagation is considered as a diffusion of a classical particle, quantum mechanics determines its effective mass $M$ and the scattering time $\tau$ entering the diffusion coefficient. Interference of different semiclassical trajectories is expected to be small in this regime characterized, as mentioned above, by the $\ell \gg \lambda$, Eq.~\eqref{applicability}. On the diagram language all contributions to $D$ where the phonon lines cross, e.g., Fig.~\ref{fig:d:diag}(b,d) contain the small parameter $\lambda/\ell$. Importantly, Eqs.~\eqref{diffusion:class} and \eqref{D:eps} disregard the contribution of the so-called self-intersecting trajectories to the diffusion coefficient, see Fig.~\ref{fig:illustr}(b). For the exciton passing the loop clockwise or counterclockwise the phase of the wavefunction $\phi_\circlearrowleft \approx \phi_\circlearrowright$ (the difference of phases arises due to a weak inelasticity of the exciton-phonon interaction), and such clock- and counterclockwise paths interfere constructively. The interference of these kind\addZakhar{s} of trajectories cannot be disregarded even if Eq.~\eqref{applicability} holds, it is because corresponding diagrams diverge in the limit of elastic scattering. As a result, the exciton propagation slows down and the diffusion coefficient acquires a quantum correction $\delta D<0$ related to the interference effects. It is the weak localization effect, being a precursor of a strong, Anderson, localization in disordered systems at $T=0$.

\begin{figure}[b]
    \centering
    \includegraphics[width=0.975\linewidth]{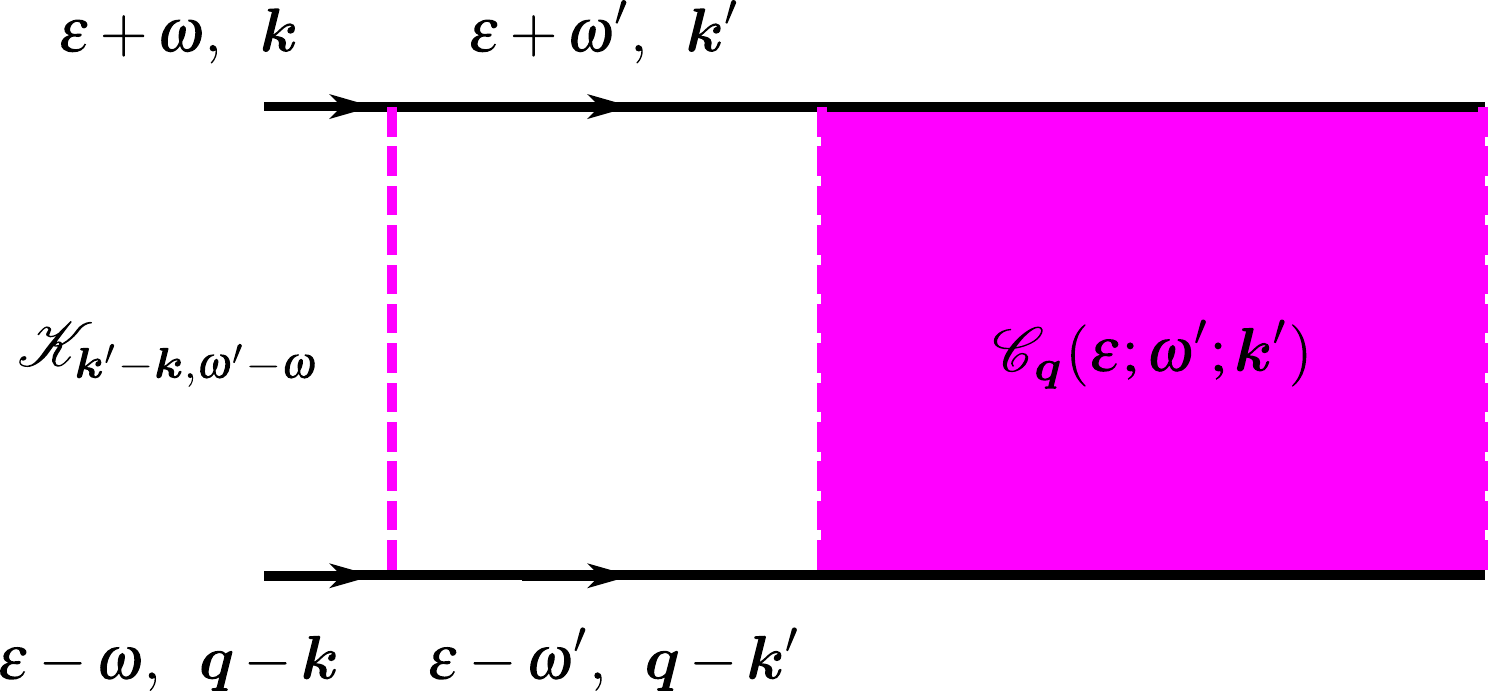}
    \caption{Kernel of integral equation~\eqref{Cooperon:Eq} for Cooperon. Solid line with arrows are the exciton Greens functions. Dashed magenta line shows the correlator of potential induced by the phonons. Magenta rectangle is the Cooperon.}
    \label{fig:Cooperon}
\end{figure}

A convenient way of evaluating $\delta D$ is to express the correction to the diffusion coefficient via the interference amplitude or Cooperon $\mathcal C_{\bm q}(\varepsilon;\omega;\bm k)$. It also describes the coherent backscattering effect and can be recast as a sum of maximally crossed diagrams, Fig.~\ref{fig:d:diag}(d).\cite{gnedin_dolginov,PhysRevLett.16.984,Gorkov:WL,1977ZhETF..72.2230I,chakravarty86,PhysRevLett.124.166802} They describe exactly the interference process presented above: The exciton undergoes the same series of the scattering events propagating back and forth along the loop. The Cooperon satisfies the integral equation with the kernel shown in Fig.~\ref{fig:Cooperon}:
\begin{multline}
    \label{Cooperon:Eq}
    \mathcal C_{\bm q}(\varepsilon;\omega;\bm k) =   \frac{M}{2\pi \hbar^3 \tau_q} \mathcal K_{\bm k_1 - \bm k,\omega_1 - \omega} \\
    + \frac{\tau_q M}{2\pi \hbar^3} \int_0^{2\pi} \frac{d\varphi}{2\pi} \int_{-\infty}^\infty \frac{d\omega'}{2\pi} \mathcal K_{\bm k' - \bm k,\omega'-\omega} \mathcal C_{\bm q}(\varepsilon;\omega';\bm k')\\
    \times \left[1 -2 \mathrm i \omega' \tau_q + \mathrm i q\ell_q \cos{\varphi'} - (q\ell_q)^2 \cos^2{\varphi'} \right].
\end{multline}
Here $\varphi$ and $\varphi'$ are the angles of the wavevectors $\bm k$ and $\bm k'$, the absolute values $k = k' = \sqrt{2M \varepsilon/\hbar^2}$ are fixed due to a weak inelasticity of the exciton-phonon scattering, and $\ell_q \equiv \ell_q(\varepsilon) = \hbar k \tau_q(\varepsilon)/M$ is the quantum mean free path; in the relevant case $q\ell_q \ll 1$. Note that the real (transport) mean free path $\ell = \hbar k \tau/M$ is determined by the momentum relaxation time $\tau$ rather than by $\tau_q$. In Eq.~\eqref{Cooperon:Eq} we omitted energy arguments of the $\tau_q$ and $\ell_q$. Cooperon $\mathcal C_{\bm q}(\varepsilon;\omega;\bm k)$ formally depends also on the final state parameters, $\bm k_1,\omega_1$, but this dependence essentially plays no role and can be omitted.

The quantity of interest for us is the angular average of $\mathcal C_{\bm q}(\varepsilon;\omega; \bm k)$. Decomposing the Cooperon into angular harmonics, taking into account that $q\ell_q \ll 1$ (that allows us to neglect the second and higher angular harmonics), and passing to the time-representation with respect to $\omega,\omega_1$, we present the equation for the reduced Cooperon
\begin{multline}
C_{\bm q}(\varepsilon;t,t_1) \\
= \tau_q^{3} \int_0^{2\pi} \frac{d\varphi}{2\pi} \int_{-\infty}^\infty\int_{-\infty}^\infty \frac{d\omega}{2\pi} \frac{d\omega_1}{2\pi} e^{-\mathrm i \omega t + \mathrm i \omega_1 t_1} \mathcal C_{\bm q}(\varepsilon;\omega;\bm k)
\end{multline}
as\footnote{In Ref.~\cite{PhysRevLett.124.166802} time variable $t$ equals to $t/2$ in the present notations.}
\begin{equation}
    \label{Cooperon:gen}
    \left[2\frac{\partial}{\partial t} + D(\varepsilon) q^2 +  \frac{\Phi(t)}{\tau_q(\varepsilon)} \right]C_{\bm q}(\varepsilon;t,t_1) = \delta(t-t_1).
\end{equation}
Here the function $\Phi(t)$ describes the dephasing of excitons; it is given by
\begin{equation}
    \label{Phi:dephasing:K}
\Phi(t) = 1 - \frac{\tau_q(\varepsilon) M}{2\pi \hbar^3} \int d\omega e^{-\mathrm i \omega t} \langle \mathcal K_{\bm k' - \bm k,\omega}\rangle,
\end{equation}
and the angular brackets denote the averaging over the scattering angle at the fixed $k$ and $k'$.
Physically, $\Phi(t) \ne 0$ is due to the fact that the scattering
events at the time-reversal related paths do not occur at exactly
the same places due to the phonon motion, see Fig.~\ref{fig:illustr}(b).

Decomposing the exponent in Eq.~\eqref{Phi:dephasing:K} up to the first nonvanishing term we obtain 
\begin{equation}
\label{Phi:ballistic}
    \Phi(t) = \frac{\delta\omega^2 t^2}{2}, \quad t\to 0,
\end{equation}
where $\hbar^2\delta \omega^2$, Eq.~\eqref{dOmega2}, is the mean squared energy transferred in the exciton-phonon scattering act.
Solving Eq.~\eqref{Cooperon:gen} we have
\begin{equation}
    \label{Cooperon:sol}
    C_{\bm q}(\varepsilon;t,t_1) = \Theta(t-t_1) \exp{\left[-\frac{D(\varepsilon) q^2 (t-t_1)}{2} -  \frac{t^3-t_1^3}{12\tau_\phi^3(\varepsilon)} \right]},
\end{equation}
Here we introduced the energy-dependent dephasing time $\tau_\phi(\varepsilon)$ 
\begin{equation}
    \label{tau:phi:phonons}
    {\tau_\phi(\varepsilon)} = \sqrt[3]{\frac{\tau_q(\varepsilon)}{\delta\omega^2}}.
\end{equation}
Taking into account that the interference contribution to the diffusion coefficient has the form\cite{PhysRevLett.124.166802} 
\begin{equation}
    \label{dD:gen}
    \delta D = - \frac{2\hbar}{Mk_B T} \sum_{\bm q} \int_0^\infty dt\int_0^\infty d\varepsilon \exp{\left( -\frac{\varepsilon}{k_B T}\right)}D(\varepsilon)C_{\bm q}(\varepsilon;t,-t),
\end{equation}
we obtain with logarithmic accuracy
\begin{equation}
    \label{dD:phonons}
    \delta D = - \frac{\hbar}{2\pi M} \ln{\left(\frac{\tau_\phi(k_B T)}{\tau(k_B T)} \right)}.
\end{equation}
Here it is assumed that the exciton lifetime  $\tau_r \gg \tau_\phi(k_B T)$, otherwise one has to replace $\tau_\phi$ by $\tau_r \tau_\phi/(\tau_r + \tau_\phi)$ in the argument of logarithm. It is also assumed that $\tau_\phi/\tau \gg 1$ and, strictly speaking, $\ln{(\tau_\phi/\tau)} \gg 1$.  Hereafter we omit the argument of $\tau_\phi$ and $\tau$ if it does not lead to a confusion.

Let us briefly address the physical interpretation of Eq.~\eqref{dD:phonons}: The efficiency of exciton interference at a closed loop is proportional to the exciton return probability to the initial point, see Fig.~\ref{fig:illustr}(b). In the case of diffusive propagation in two dimensions it is given by\cite{Gorkov:WL,chakravarty86,PhysRevLett.58.2106,PhysRevLett.124.166802} $\int_{t_{\rm min}}^{t_{\rm max}} dt/t = \ln{(t_{\rm max}/t_{\rm min})}$ where the lower limit corresponds to the shortest propagation time $t_{\rm min} \sim \tau$, while the upper limit corresponds to the maximum time where the coherence of excitons is preserved, $\tau_{\rm max} \sim \min\{\tau_\phi,\tau_r\}$. If the disorder was static and excitons had infinite lifetime, then $\tau_\phi \to \infty$ and the return probability diverges. As a result, $\delta D$ formally becomes infinite (and negative) meaning that interference effects lead to a strong, Anderson, localization of the quasiparticles.\cite{RevModPhys.57.287,RevModPhys.80.1355} The localization length, however, can be extremely large $l_{\rm loc} \sim \ell \exp{(\ell/\lambda)}$ and exceed the size of the sample. It allows us to consider excitons at $\lambda/\ell \ll 1$, Eq.~\eqref{applicability}\addZakhar{,} as almost free.

Equations~\eqref{tau:phi:phonons} and \eqref{dD:phonons} are the main results of this section. In the next section we study the temperature dependence of the diffusion coefficient for various regimes of exciton-phonon interaction.

\section{Results}

In order to determine the exciton diffusion coefficient $D=D_{cl} + \delta D$ we need to find the momentum relaxation time $\tau(\varepsilon)$ and phase relaxation time $\tau_\phi(\varepsilon)$. Then, using Eq.~\eqref{diffusion:class} we can readily evaluate the semiclassical value of the diffusivity, $D_{cl}$, and using Eq.~\eqref{dD:phonons} the interference correction, $\delta D$. We are mainly interested in the temperature dependence of $D_{cl}$ and $\delta D$ for different types of involved phonons and, accordingly, different mechanisms of exciton-phonon interaction. Depending on the phonon dispersion and specific coupling mechanism the correlation function $\mathcal K_{\bm q,\omega}$ differs resulting in different temperature dependence of the exciton diffusion coefficient.

\subsection{Longitudinal acoustic phonon scattering}\label{sec:LA}

Let us start with a conventional situation where excitons interact with longitudinal acoustic phonons with linear dispersion 
\begin{equation}
\label{LA}
    \Omega_q = s q, \quad \mbox{LA phonon,}
\end{equation} 
with $s$ being the speed of sound. It is a typical case realized in monolayer semiconductors. Quantum corrections to the exciton diffusion coefficient in this regime have been addressed in Refs.~\onlinecite{PhysRevLett.124.166802,PhysRevLett.127.076801}. The exciton-phonon interaction matrix elements read [see, e.g., Ref.~\onlinecite{shree2018exciton}]
\begin{equation}
\label{matrix:elements}
U(\bm r, t) = \sum_{\bm q} \sqrt{\frac{\hbar}{2\rho \Omega_q}} \mathrm i q \Xi \left( b_{\bm q}  e^{\mathrm i \bm q \bm r- \mathrm i \Omega_q t}- b_{\bm q}^\dag  e^{-\mathrm i \bm q \bm r+\mathrm i \Omega_q t}\right),
\end{equation}
where $\bm q$ is the phonon wavevector,  $\Xi$ is the exciton deformation potentials (being the difference of the conduction and valence band deformation potentials), $b_{\bm q}^\dag$ ($b_{\bm q}$) are the phonon creation (annihilation) operators; as before, the normalization area is set to unity. Neglecting the phonon damping we arrive at the following expression for the Fourier transform of the  correlation function of the phonon-induced potential
\begin{equation}
\label{corr:UU:F:1}
\mathcal K_{\bm q,\omega} = \frac{\pi  k_B T}{\rho s^2} \Xi^2 
[\delta(\omega - \Omega_q) +  \delta(\omega + \Omega_q) ].
\end{equation}
Equation~\eqref{corr:UU:F:1} is applicable for sufficiently high temperatures where the phonon occupancy reads $k_B T/\hbar\Omega_q \gg 1$, it is exactly the temperature range of interest. Using Eq.~\eqref{times:correlator:gen} 
we obtain
\begin{equation}
\label{momentum:1}
\frac{1}{\tau(\varepsilon)} =  \frac{1}{\tau_q(\varepsilon)} =\frac{k_B T}{Ms^2} \frac{1}{\tau_0}, \quad \tau_0^{-1} = {M^2\Xi^2}/{(\rho \hbar^3)}.
\end{equation}
It is noteworthy that the correlation function of the phonon-induced potential $\mathcal K_{\bm q, \omega}$ does not explicitly depend on the phonon wavevector $\bm q$, as a result momentum and quantum relaxation times are the same in this mechanism.
Equation~\eqref{momentum:1} is valid for reasonably high temperatures $T \gtrsim Ms^2/k_B \sim 1$~K, otherwise the exciton-phonon interaction is strongly inelastic and the quantum infererence effects are negligible. Note that the momentum relaxation time is energy independent in this mechanism. It depends on the temperature via the phonon occupancies. As a result, substituting $\tau(\varepsilon)$ from Eq.~\eqref{momentum:1} into Eq.~\eqref{diffusion:class} we have for the semiclassical value of the diffusion coefficient  
\begin{equation}
\label{D:ac}
D_{cl}= \frac{k_B T}{M} \tau(k_B T) = s^2 \tau_0.
\end{equation}
In this case it is temperature-independent.
 
The quantum correction to the diffusion coefficient is determined, according to Eq.~\eqref{dD:phonons},  by the phase relaxation time $\tau_\phi$. It follows from Eq.~\eqref{tau:phi:phonons} that\footnote{A numerical coefficient is omitted, it plays no role in the argument of the logarithm.}
\begin{equation}
    \label{tau:phi:ac}
\tau_\phi = \left(\frac{\hbar^2 \tau_0}{(k_BT)^2}\right)^{1/3} \propto T^{-2/3}.
\end{equation}
As discussed in Ref.~\onlinecite{PhysRevLett.124.166802} in this case, $\tau_\phi/\tau \propto T^{1/3}$ and $\delta D$ decreases with increasing the temperature (its absolute value increases). For sufficiently long $\tau_\phi$ the increase of $|\delta D|$ saturates due to the fact that $\tau_\phi$ becomes comparable with $\tau_r$. Also, with the temperature increase, almost dispersionless optical phonons and zone-edge acoustic phonons start to play a role resulting in rapid dephasing with $\tau_\phi\sim \tau$. Estimates show that in the temperature range from $1\ldots 3$~K up to $20\ldots 30$~K the diffusion coefficient can be controlled by the weak localization effect. The experimental signatures of non-classical exciton propagation have been recently reported.\cite{PhysRevLett.127.076801}

\subsection{Flexural phonon scattering in two-layer TMDCs}\label{sec:flex:2}

We now turn to the regime where excitons mainly interact with flexural phonons. Such situation has not been studied from the viewpoint of the exciton transport to the best of our knowledge. The flexural modes with the out-of-plane atomic displacements are much softer than the longitudinal acoustic modes.\cite{Katsnelson2020}  Accordingly, we take the dispersion in the form
\begin{equation}
    \label{flex}
    \Omega_q = \omega_0 + \varkappa q^2, \quad  \mbox{flexural phonon,}
\end{equation} 
where the ``stiffness'' $\varkappa$  is related to the bending rigidity of the system and $\omega_0$ is the cut-off frequency. We disregard the effects of anharmonic interaction between the in-plane and out-of-plane phonons that renormalize bending rigidity and phonon dispersion.\cite{Le-Doussal:2018vk,Gornyi:2016aa,PhysRevB.94.195430} 

Efficient interaction of excitons with the flexural phonons can be realized, e.g., in bilayer TMDC where the electron and hole forming the exciton are localized in different layers and the phonon mode in question is an antisymmetric mode of layer vibrations where the layers move towards each other and back. In such a case the coupling of excitons with the flexural vibrations is particularly pronounced since the phonon induced variation of the interlayer distance directly corresponds to a variation of the electron-hole Coulomb interaction (the exciton-flexural phonon coupling in monolayers is briefly discussed below). It is described by the following potential\cite{https://doi.org/10.1002/andp.202000339,2022arXiv220212143I}
\begin{equation}
\label{matrix:elements:flex}
U(\bm r, t) = \sum_{\bm q} \sqrt{\frac{\hbar}{2\rho \Omega_q}} \Xi' b_{\bm q}  e^{\mathrm i \bm q \bm r - \mathrm i \Omega_q t}+{\rm c.c.},
\end{equation}
where $\Xi'$ is the corresponding interlayer deformation potential parameter, $\Xi'\sim e^2/\mbox{max}\{L^2,a_B^2\}$ with $L$ being the interlayer distance and $a_B$ the two-dimensional exciton Bohr radius.\footnote{The coupling with flexural phonons can be relatively strong for two reasons: (i) due to soft dispersion the quantum of the flexural phonon displacement is typically larger than for the longitudinal acoustic one and (ii) effective interaction constant $\Xi'$ can be on the order of 100~meV/nm being comparable or even higher than $q\Xi$ for the LA mode.} We note that it is a simplified form of the exciton-phonon interaction in bilayer systems, which for example does not include possible mixing of the spatially direct and indirect excitons.\cite{Barr2022}
 Hereafter we assume that the weak coupling regime is fulfilled, thus polaron effects studied in Refs.~\onlinecite{https://doi.org/10.1002/andp.202000339,2022arXiv220212143I} are relatively weak. Instead of Eq.~\eqref{corr:UU:F:1} we now have the correlator of the phonon-induced potential in the form
\begin{multline}
\label{corr:UU:F:1:flex}
\mathcal K_{\bm q,\omega} = \frac{\pi  k_B T }{\rho \Omega_q^2} \Xi'^2 
[\delta(\omega - \Omega_q) +  \delta(\omega + \Omega_q) ].
\end{multline}
Making use of Eq.~\eqref{momentum} we have in this case
\begin{equation}
    \label{tau:flex}
    \frac{1}{\tau(\varepsilon)} = \frac{k_B T M}{\rho \hbar^3} \frac{\Xi'^2}{\omega_0^2} \frac{w_0^3}{\left[{(2+w_0)w_0}\right]^{3/2}} .
\end{equation}
Here the parameter $$w_0 \equiv w_0(\varepsilon) = \frac{\omega_0 \hbar^2}{4M\varepsilon\varkappa}$$ determines the role of the phonon dispersion: at small $w_0 \ll 1$ the phonon energy corresponding to the transferred wavevector $\sim \varkappa q^2 \gg\omega_0$, i.e., the phonon dispersion is significant, while at large $w_0 \gg 1$ the phonon dispersion plays no role and $\Omega_q$ can be replaced by a cut-off value $\omega_0$. The quantum scattering time is readily evaluated using Eq.~\eqref{quantum} and reads
\begin{equation}
    \label{tau:q:flex}
    \frac{1}{\tau_q(\varepsilon)} = \frac{k_B T M}{\rho \hbar^3} \frac{\Xi'^2}{\omega_0^2} \frac{w_0^2(1+w_0)}{\left[{(2+w_0)w_0}\right]^{3/2}} .
\end{equation}
Making use of Eq.~\eqref{tau:phi:phonons} we obtain
\begin{equation}
\label{tau:phi:flex:gen}
\tau_\phi = \left(\frac{\rho\hbar^3}{k_B T M \Xi'^2}\right)^{1/3} \propto T^{-1/3}.
\end{equation}

\begin{figure}[t]
    \centering
    \includegraphics[width=0.97\linewidth]{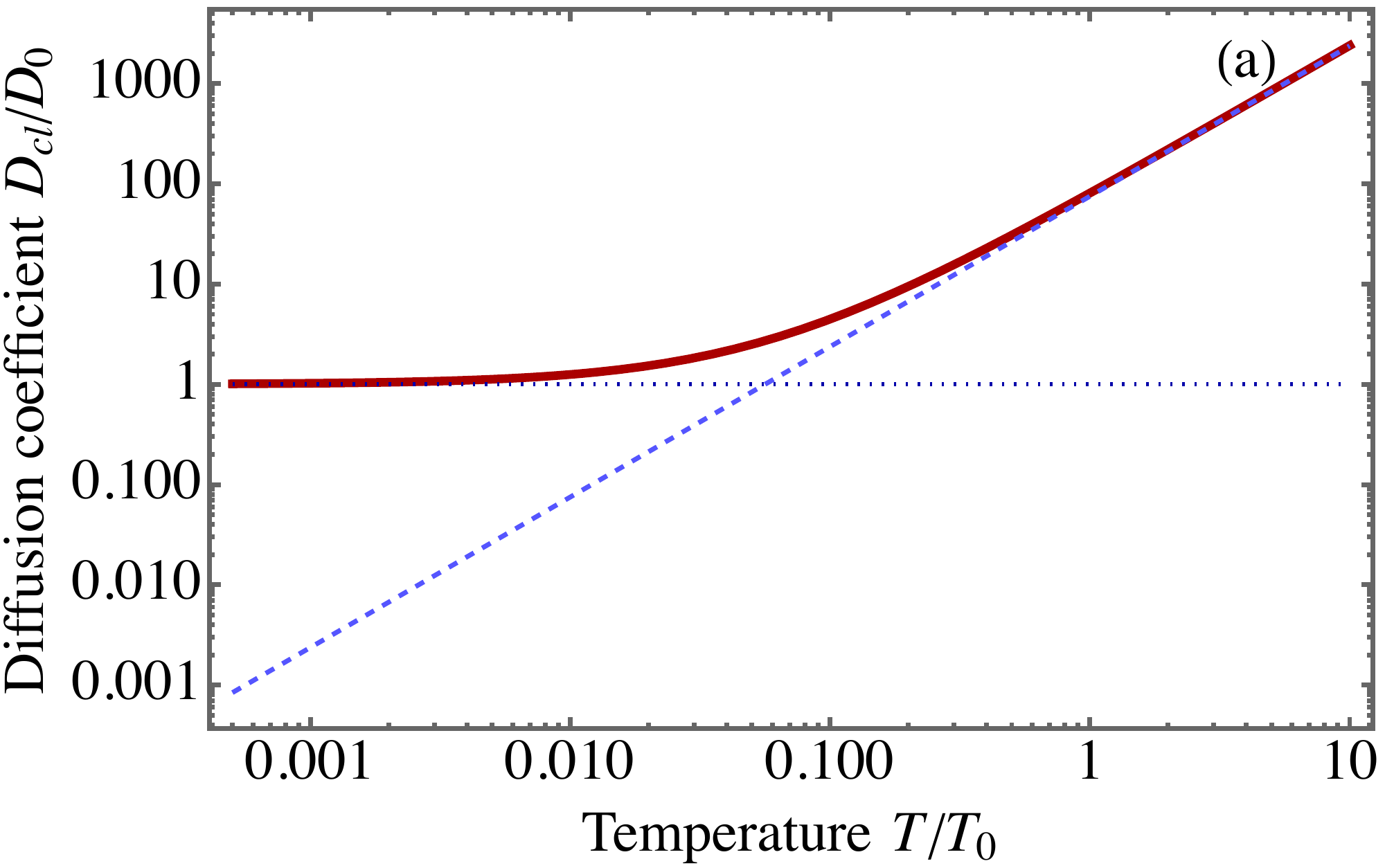}
    \includegraphics[width=0.97\linewidth]{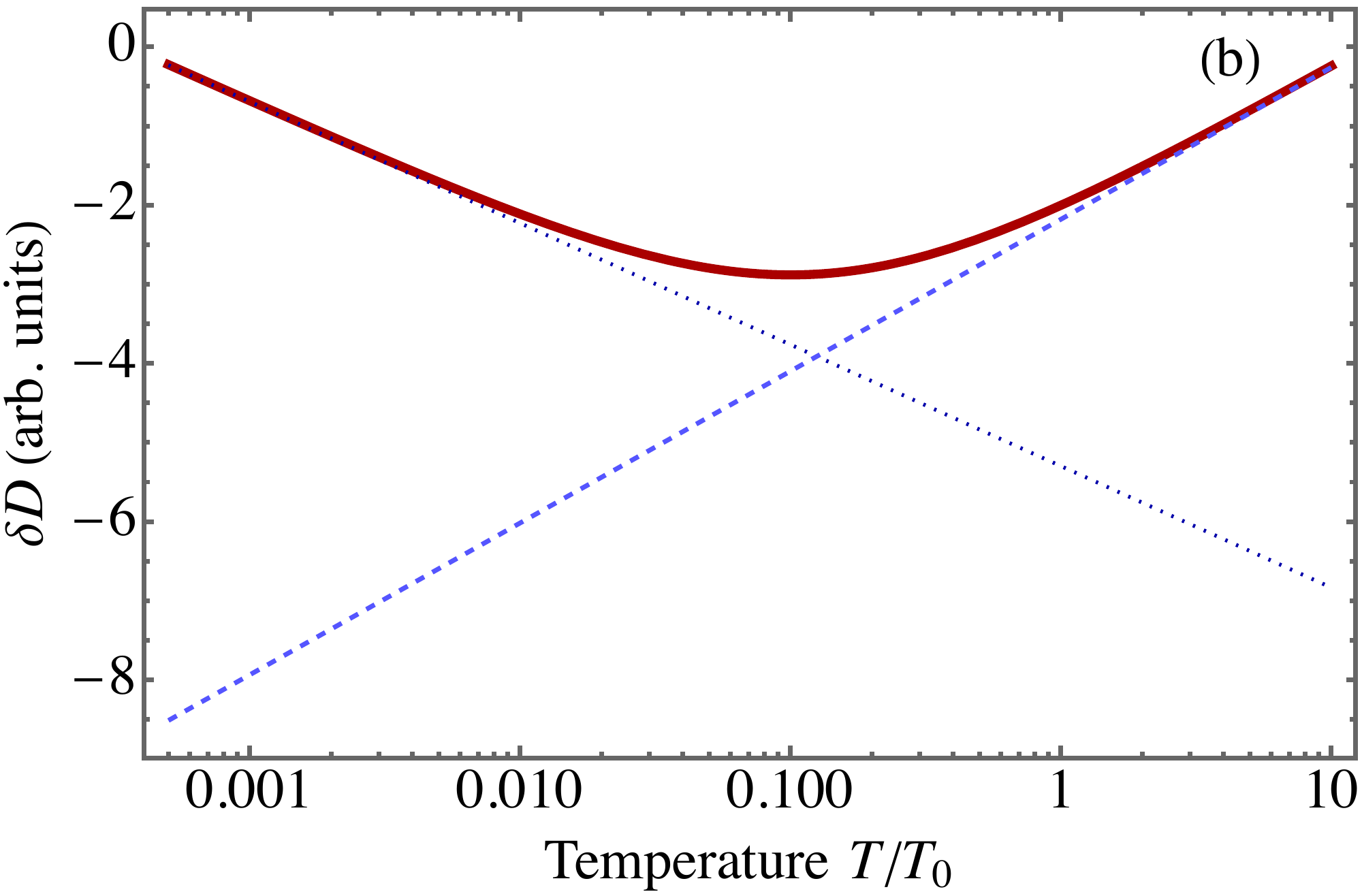}
    \caption{Diffusion coefficient at the exciton-flexural phonon scattering. (a) Classical value $D_{cl}$ calculated after Eqs.~\eqref{diffusion:class} and \eqref{tau:flex} as \addSivan{a} function of temperature (red solid line), dotted and dashed lines show the low- and high-temperature asymptotics, Eqs.~\eqref{D:flex:large} and \eqref{D:flex:small}, respectively. (b) Quantum correction, $\delta D$. Red solid line presents the result of \addSivan{the} calculation using Eqs.~\eqref{dD:phonons}, \eqref{tau:flex} and \eqref{tau:phi:flex:gen}. Dotted and dashed lines present the  low- and high-temperature asymptotics. $D_0=\rho\hbar^3 \omega_0^2/(M^2\Xi'^2)$, $T_0 = \omega_0\hbar^2/(k_B M\varkappa)$.}
    \label{fig:D:flex}
\end{figure}

Let us consider two important limits. At $w_0(k_B T) \gg 1$ the phonon dispersion is unimportant and $\Omega_q \approx \omega_0$. In this regime
\begin{equation}
    \label{tau:flex:large}
    \frac{1}{\tau(\varepsilon)} =  \frac{1}{\tau_q(\varepsilon)} = \frac{k_B T M}{\rho \hbar^3} \frac{\Xi'^2}{\omega_0^2},
\end{equation}
hence the momentum relaxation time is energy independent and inversely proportional to the temperature. For the semiclassical diffusion coefficient we have
\begin{equation}
    \label{D:flex:large}
    D_{cl} = \frac{\rho\hbar^3\omega_0^2}{M^2\Xi'^2}.
\end{equation}
It is temperature independent similarly to the case of excitons interacting with the longitudinal acoustic phonons studied above. For $\hbar\omega_0 \gtrsim k_B T$ the exciton-phonon scattering is strongly inelastic, $\tau_\phi \sim \tau$, and the quantum contribution $\delta D$ is negligible; we do not consider this regime. By constrast, at $\hbar\omega_0 \ll k_B T$ the exciton-phonon interaction is quasi-elastic and according to Eqs.~\eqref{tau:phi:flex:gen}
\begin{equation}
        \label{tau:phi:flex:large}
\tau_\phi = \left( \frac{\tau}{\omega_0^2}\right)^{1/3}.
\end{equation}
The ratio $\tau_\phi/\tau \propto (\omega_0\tau)^{-2/3} \propto T^{2/3}$. Thus, the temperature dependence of interference contribution to the diffusion coefficient $\delta D$ is stronger in this case as compared to the situation of longitudinal acoustic phonon scattering. Like in the case considered above the $|\delta D|$ increases with increase in the temperature until the parameter $w_0(k_B T)$ becomes small, see below, or strongly inelastic scattering processes come into play. Similar behavior of the diffusion coefficient is expected if the excitons interact with dispersionless phonons (e.g., optical) at sufficiently high temperatures where the quasi-elasticity is realized, $k_B T \gg \hbar\omega_0$.

In the opposite regime corresponding to high-temperatures where $w_0(k_B T)\to 0$, 
\begin{equation}
    \label{tau:flex:small}
    \frac{1}{\tau(\varepsilon)} = \frac{k_B T M}{\rho \hbar^3} \frac{\Xi'^2}{\sqrt{\omega_0}} \left(\frac{\hbar^2}{8 M\varepsilon\varkappa}\right)^{3/2},
\end{equation}
and the momentums scattering time scales as $\tau(\varepsilon) \propto \varepsilon^{3/2}T^{-1}$. 
In this regime the exciton-phonon scattering is strongly anisotropic with $\tau(\varepsilon) \ll \tau_q(\varepsilon)$.
Making use of Eqs.~\eqref{diffusion:class} and \eqref{D:eps} we obtain for the diffusion coefficient
\begin{equation}
\label{D:flex:small}
D_{cl} = 30\sqrt{2\pi} \frac{\rho\hbar^3\omega_0^2}{M^2\Xi'^2} \left(\frac{M k_B T\varkappa}{\omega_0\hbar^2} \right)^{3/2}.
\end{equation}
Here $D\propto T^{3/2}$ as results from $\varepsilon^{3/2}/T$ dependence of the scattering time in Eq.~\eqref{tau:flex:small}. It follows from Eq.~\eqref{tau:phi:flex:gen} that $\tau_\phi(k_B T)/\tau(k_B T)\propto T^{-5/6}$. In this case the absolute value of the quantum correction to the diffusion coefficient $|\delta D|$ decreases with increasing the temperature.

These results are illustrated in Fig.~\ref{fig:D:flex} where the panel (a) demonstrates the classical value of the diffusion coefficient and the bottom panel shows the quantum correction. Solid curves are calculated using the general expressions \eqref{diffusion:class}, \eqref{tau:flex}, and \eqref{tau:phi:flex:gen}, while dotted and dashed curves present the asymptotics at relatively low and high temperatures, $T \ll T_0$ and $T \gg T_0$, respectively, where $k_B T_0 = \omega_0\hbar^2/(M\varkappa)$ determines the crossover between the regimes where the flexural phonon dispersion is relevant (high temperatures) or not (low temperatures). Overall behavior of the calculated diffusion coefficient is consistent with the analysis of the limits above. The semiclassical $D_{cl}$ monotonously increases with the temperature increase, because the flexural phonon scattering becomes less efficient at high temperatures, while $\delta D$ demonstrates non-monotonous behavior: It first decreases and then increases with the temperature increase.

\subsection{Flexural phonon scattering in monolayers}\label{sec:flex:2ph}

Next, let us briefly address the effects of exciton scattering by flexural phonons in monolayer semiconductors. In this regime the exciton-phonon interaction is suppressed: Indeed, as symmetry analysis shows the out-of-plane vibrations of atoms do not affect quasiparticles in a monolayer in the first order in the displacements. In the second order, the out-of-plane phonons produce the potential in the form~\cite{M.IKatsnelson01282008,PhysRevB.100.075417}
\begin{equation}
\label{flex:ML:2ph}
U(\bm r,t) = \frac{\Xi}{2} \left[ \left(\frac{\partial \zeta(\bm r,t)}{\partial x}\right)^2+\left(\frac{\partial \zeta(\bm r,t)}{\partial y}\right)^2\right],
\end{equation}
where $\zeta(\bm r,t)$ are the out-of-plane lattice displacements. Equation~\eqref{flex:ML:2ph} shows that the two-phonon processes are relevant in this case. We evaluate the correlation function extending the approach of Ref.~\onlinecite{PhysRevB.100.075417} where the same-time correlation function has been evaluated. After some algebra we obtain
\begin{equation}
\label{K:flex:ML:0}
    \mathcal K_{\bm q, \omega} = \frac{(1-\nu)^2}{32} \Xi^2 \frac{(k_BT)^2}{\rho^2\varkappa^5q^4}F\left(\frac{\varkappa q^2}{\omega}\right),
\end{equation}
where 
\begin{equation}
\label{fun:F}
F(x) = \begin{cases}
1 \quad, \quad  0<|x|<1/2\\
3 - \frac3{2}|x|^2 + \frac1{2}|x|^3  \quad, \quad 1/2 < |x| < 1,\\
\frac3{2}|x|^2+\frac1{2}|x|^3  \quad, \quad  1<|x|.
\end{cases}
\end{equation}
and $\nu$ is 2D Poisson coefficient. Introducing renormalized deformation potential 
\[
\tilde \Xi^2 = \frac{3(1-\nu)^2}{32\pi}\Xi^2, 
\]
we obtain the frequency-integrated correlator of the phonon-induced potential takes (at not too low temperatures and too small wavevectors) the form~\cite{PhysRevB.100.075417}
\begin{equation}
    \label{K:flex:ML}
    \int_{-\infty}^\infty d\omega \mathcal K_{\bm q,\omega} = {2\pi}\frac{(k_B T)^2}{\rho^2\varkappa^4 q^2}\tilde \Xi^2,
\end{equation}
Combining Eqs.~\eqref{rates:K:alt} and ~\eqref{K:flex:ML} we obtain
\begin{subequations}
    \begin{equation}
        \label{tau:flex:ML}
        \frac{1}{\tau(\varepsilon)} = \frac{(k_B T)^2\tilde \Xi^2}{4\varepsilon\varkappa^4\rho^2\hbar}
    \end{equation}
        \begin{equation}
        \label{tau:q:flex:ML}
        \frac{1}{\tau_q(\varepsilon)} = \frac{1}{\tau(\varepsilon)} \sqrt{\frac{2M \varepsilon}{q_*^2 \hbar^2}},
    \end{equation}
\end{subequations}
where $q_*$ is the cut-off wavevector, at $q\lesssim q_*$ anharmonic coupling between the modes becomes important.\cite{Le-Doussal:2018vk,Gornyi:2016aa,PhysRevB.94.195430,Katsnelson2020} Correspondingly, the semiclassical value of the diffusion coefficient reads
\begin{equation}
    \label{D:cl:flex:ML}
    D_{cl} = \frac{8\rho^2\hbar^2\varkappa^4}{M\tilde\Xi^2},
\end{equation}
and $D_{cl}$ is temperature-independent.

In order to calculate the quantum correction we need to evaluate the dephasing time. To that end, we have to take into account inelasticity of the exciton-phonon interaction. It follows from Eqs.~\eqref{K:flex:ML:0} and \eqref{fun:F} that at $\omega \to \infty$ the correlation function $\mathcal K_{\bm q,\omega} \propto \omega^{-2}$. Consequently the mean squared energy transferred in the exciton-two-phonon interaction process, Eq.~\eqref{dOmega2}, diverges. To overcome this problem, we examine in more detail the temporal dependence of the dephasing function $\Phi(t)$ introduced in Eq.~\eqref{Phi:dephasing:K}. Performing the Fourier transform of Eq.~\eqref{K:flex:ML:0} and decomposing it in series at $t\to 0$ we have
\begin{equation}
    \label{Phi:flex:2}
    \Phi \approx \pi \frac{\tau_q(\varepsilon)}{\tau(\varepsilon)}  \frac{M\varepsilon}{\hbar^2}\varkappa |t|  + \ldots =\frac{\pi}{2}\varkappa q_* \sqrt{\frac{2M\varepsilon}{\hbar^2}} |t| + \ldots,
\end{equation}
where $t^2$ and higher-order terms are omitted. Hence, the dephasing is different as compared to Eq.~\eqref{Cooperon:sol}. Such form of dephasing is also typical for the case where the scattering is caused by the slowly moving particles.\cite{PhysRevA.36.5729}  Solving Eq.~\eqref{Cooperon:gen} with $\Phi(t)$ in the form of Eq.~\eqref{Phi:flex:2} we obtain the factor in the Cooperon $C_q(\varepsilon;t,t_1)$ responsible for the dephasing in the form 
\begin{equation}
\label{Cooperon:flex:2}
C_q(\varepsilon;t,t_1) \propto \exp{\left[-\frac{t^2-t_1^2}{4\tau_\phi^2(\varepsilon)} \right]},
\end{equation}
with 
\begin{equation}
\label{tau:phi:flex:2}
\tau_\phi(\varepsilon) = \sqrt{\frac{\hbar^2 \tau(\varepsilon)}{\pi M\varepsilon \varkappa}}.
\end{equation}
Thus, $\tau_\phi(k_B T) \propto 1/(k_B T)$ and the ratio $\tau_\phi(k_B T)/\tau(k_B T)$ does not depend on the temperature resulting in a constant $\delta D$.

\subsection{Exciton interaction with damped phonons}

Another interesting regime of the exciton weak localization may occur if the phonon damping caused, e.g., by the defects and anharmonicity, is strong. For completeness, we analyze this scenario as well. For definiteness, we focus on the excitons interacting with longitudinal acoustic phonons. The most straightforward way to allow for the phonon damping is to replace the $\delta$-functions in $\mathcal K_{\bm q,\omega}$, Eq.~\eqref{corr:UU:F:1}, by Lorentzians
\begin{equation}
\label{Lorentz:1}
\delta(\omega \pm \Omega_q) \to \frac{1}{\pi} \frac{\gamma}{(\omega\pm\Omega_q)^2+\gamma^2},
\end{equation}  
where $\gamma$ is the damping of the phonon mode. Such replacement does not affect the values of the scattering times $\tau_q(\varepsilon)$ and $\tau_q$, because these quantities are determined by the frequency-integrated correlator. However, under the replacement~\eqref{Lorentz:1} the dephasing becomes quite specific. Indeed, for the dephasing function $\Phi(t)$ in Eq.~\eqref{Phi:dephasing:K} instead of Eq.~\eqref{Phi:ballistic} we obtain
\begin{equation}
\label{Phi:Lorentz:1}
\Phi(t) = 1- e^{-\gamma|t|}\langle\cos{(\Omega_qt)}\rangle \approx \gamma|t| - \frac{\langle \Omega_q^2\rangle t^2}{2}, \quad t\to 0,
\end{equation}
similarly to the situation studied above in Sec.~\ref{sec:flex:2ph}, see Eq.~\eqref{Phi:flex:2}.
As a result, for sufficiently strong phonon damping, the factor in the Cooperon $C_q(\varepsilon;t,t_1)$ responsible for the dephasing reads [cf. Eq.~\eqref{Cooperon:flex:2}]
\begin{equation}
\label{Cooperon:2}
C_q(\varepsilon;t,t_1) \propto \exp{\left[-\frac{t^2-t_1^2}{4\tau_\phi^2(\varepsilon)} \right]},
\end{equation}
with 
\begin{equation}
\label{tau:phi:Lorentz:1}
\tau_\phi(\varepsilon) = \sqrt{\frac{\tau_q(\varepsilon)}{\gamma}}.
\end{equation}
Note that in this case the dephasing is controlled by the phonon damping and $\tau_\phi/\tau \propto T^{1/2}$ for the longitudinal acoustic phonon scattering provided that $\gamma$ is temperature independent.

However, the replacement~\eqref{Lorentz:1} may not be entirely appropriate for the phonons that obey mechanical equations of motion. The general expression for the phonon-induced correlator can be obtained by the following replacement\footnote{It can be derived using the Greens function method or via the mechanical equations of motion with random Langevin forces taking into account that the equipartition theorem is valid for a damped harmonic oscillator.}
\begin{multline}
\label{Lorentz:2}
\delta(\omega + \Omega_q) + \delta(\omega - \Omega_q) \to \frac{1}{\pi} \frac{4\gamma\Omega_q^2}{(\omega^2-\Omega_q^2)^2+4\gamma^2\omega^2}\\
= \frac{\mathrm i }{\pi \omega} \left(\frac{\Omega_q^2}{\omega^2 - \Omega_q^2 +2\mathrm i \gamma\omega} - \frac{\Omega_q^2}{\omega^2 - \Omega_q^2 - 2\mathrm i \gamma\omega} \right).
\end{multline}
In such case, the dephasing  function $\Phi(t)$ in Eq.~\eqref{Phi:dephasing:K} takes the form
\begin{equation}
\label{Phi:Lorentz:2}
\Phi(t) = 1- e^{-\gamma|t|}\left\langle\cos{\tilde\Omega_q t} + \frac{\gamma}{\tilde\Omega_q} \sin{\tilde \Omega_q |t|} \right\rangle,
\end{equation}  
where $\tilde \Omega_q = \sqrt{\Omega_q^2 - \gamma^2}$. Equation~\eqref{Phi:Lorentz:2} is valid for any relation between $\gamma$ and $\Omega_q$. We abstrain from the detailed analysis of all phonon damping regimes and corresponding electron dephasing described by $\Phi(t)$ in Eq.~\eqref{Phi:Lorentz:2}. Here we emphasis that if $\gamma$ is sufficiently large then the dephasing time is given by Eq.~\eqref{tau:phi:Lorentz:1} derived using a simplified form of the correlation function $\mathcal K_{\bm q,\omega}$. 

\subsection{Effect of free charge carriers on exciton weak localization}

To conclude this part, we discuss the effect of the free carriers on the exciton weak localization. To that end we note that the electron and hole effective masses are comparable in 2D TMDC crystals~\cite{2053-1583-2-2-022001,Durnev_2018,RevModPhys.90.021001}. Thus, exciton-electron or exciton-hole scattering is highly inelastic since the masses of the exciton and charge carrier differ by a factor $\sim 2$. Thus, if exciton diffusion is controlled by the exciton-free carrier scattering, then the dephasing time due to the electron-exciton interaction $\tau_\phi^{eX} = \tau^{eX}$ where $\tau^{eX}$ is the exciton-charge carrier momentum relaxation time, and quantum correction to the diffusion coefficient is absent. 

An interesting regime, however, may occur at elevated temperatures and relatively low electron densities where the following conditions are met:
\begin{equation}
    \tau^{ph} \ll \tau^{eX} \ll \tau_\phi^{ph}.
\end{equation}
Here we use the subscript $ph$ to denote the times relevant for the exciton-phonon scattering and we omitted the arguments $k_B T$ for brevity.
In this case, the semiclassical value of the exciton diffusion coefficient is controlled by the exciton-longitudinal acoustic phonon interaction, while the dephasing mainly occurs due to the exciton-electron (or exciton-hole) scattering. For not too high carrier densities $n$ we can estimate for the exciton-free carrier scattering rate as\cite{wagner:trions}
\begin{equation}
\label{tau:e:X}
\frac{1}{\tau^{eX}} \approx  \frac{\hbar^2 n}{M} \frac{3\pi^2}{\ln^2[\hbar/(2\tau^{ph} E_{b,T})] + \pi^2/4},
\end{equation}
where $E_{b,T}$ is the trion binding energy. In this case the dephasing rate linearly depends on the charge carrier density $n$ and only weakly on the temperature, via the logarithm in the denominator of Eq.~\eqref{tau:e:X}. Hence, $\tau_\phi/\tau \propto T/n$ resulting in the increase of the quantum correction absolute value $|\delta D|$ with increase of the temperature and decrease of the free carrier density.

\section{Discussion and conclusion}

An overview of the exciton transport regimes is given in Table~\ref{tab:summary}. It is seen that in several cases the semiclassical value of the exciton diffusion coefficient is temperature independent. In these regimes the quantum interference of excitons -- the weak localization effect -- becomes particularly important: It controls the temperature dependence of the diffusion coefficient $D = D_{cl} + \delta D$. Since for all discussed scattering regimes the exciton-phonon scattering is quasi-elastic the dephasing time can be sufficiently long, $\tau_\phi \gg \tau$ and the $\delta D \propto \ln{\tau_\phi/\tau}$ can be sizeable. The absolute value of $\delta D$ is controlled by the product of the relatively large temperature-dependent dimensionless logarithmic factor, $\ln{\tau_\phi/\tau}$, and the system-dependent dimensional constant, $\hbar/(2\pi M)$, the latter being on the order of $0.1\ldots 0.5$~cm$^2$/s depending on the exciton effective mass.

\begin{table}[t]
\caption{\label{tab:summary} Summary of exciton transport regimes. $\downarrow$ denotes the decrease of $\delta D$ with increase of the temperature ($|\delta D|$ increases), $\uparrow$ denotes the increase of $\delta D$ with increase of the temperature ($|\delta D|$ decreases).}
\begin{tabular}{c||c|c|c}
Scattering mechanism & $D_{cl}(T)$ & $\tau_\phi/\tau$ & $\delta D(T)<0$\\
\hline\hline
LA & $T^0$, Eq.~\eqref{D:ac} & $T^{1/3}$ & $\downarrow$\\
flexural, 1ML & $T^0$, Eq.~\eqref{D:cl:flex:ML} & $T^{0}$ & const\\
flexural, 2ML, $T<T_0$ & $T^0$, Eq.~\eqref{D:flex:large} & $T^{2/3}$ & $\downarrow$\\
flexural, 2ML, $T>T_0$ & $T^{3/2}$, Eq.~\eqref{D:flex:large} & $T^{-5/6}$ & $\uparrow$\\
LA, overdamped & $T^0$, Eq.~\eqref{D:ac} & $T^{1/2}$ & $\downarrow$\\
LA + free $e/h$ & $T^0$, Eq.~\eqref{D:ac} & $T/n$ & $\downarrow$
\end{tabular}
\end{table}

The scenario of exciton scattering by the longitudinal acoustic phonons with linear dispersion is the most relevant for monolayer and bilayer TMDCs at low temperatures. The signatures of exciton weak localization in exciton transport in WSe$_2$ ML have been reported in Ref.~\onlinecite{PhysRevLett.127.076801}: In the temperature range where the $D_{cl}$ was expected to remain temperature independent the observed diffusion coefficient decreased with increasing the temperature. Additional evidence for nonclassical exciton propagation comes from the temperature-induced exciton line broadening. The latter is determined by a single particle scattering time, $\tau_q(\varepsilon)$. The temperature dependence of linewidth in Ref.~\onlinecite{PhysRevLett.127.076801} fully corresponds to the acoustic phonon scattering where $\tau_q^{-1} \equiv \tau^{-1} \propto k_B T$. 

Still, the magnitude of the effect observed experimentally was larger than the theoretical prediction. The discrepancy between the experiment and the theory can be related to the fact that the present model is simplified: (i) it neglects specifics of the exciton and phonon bandstructure in 2D TMDCs and (ii) the applicability condition~\eqref{applicability} is not, strictly speaking, fulfilled. In respect to the point (i) we note that the development of the exciton weak localization theory with allowance for all features of the TMDCs electronic and vibrational spectra is an interesting problem beyond the scope of the present paper. 

As for the second point, namely, violation of Eq.~\eqref{applicability}, which is indeed the case in the experiment (see also estimates in Ref.~\onlinecite{PhysRevLett.124.166802}), two interrelated phenomena are possible. On the one hand, if $|\delta D|$ becomes comparable with $D$, then the first-order quantum corrections described by the model above are insufficient to calculate the diffusion coefficient. One can attempt to capture the higher-order effects described by more sophisticated diagrams like in Fig.~\ref{fig:d:diag}(b) within the so-called self-consistent theory of localization,\cite{PhysRevLett.45.842,PhysRevB.22.4666,sadovskii:1982:eng} although the approximations in this approach are hardly controllable. The self-consistent theory of localization leads to the following equation for the total diffusion coefficient $D$:
\begin{equation}
\label{self-consistent}
\frac{D}{D_{cl}} = 1 - \frac{\hbar}{4\pi M D_{cl}} \ln{\left(\frac{D\tau_\phi}{D_{cl}\tau} \right)}.
\end{equation}
It can be readily solved via the Lambert $\mathcal W$ function (product logarithm) as
\begin{equation}
    \label{D:self}
    D = \frac{1}{4\pi}\mathcal W\left(\frac{\tau}{\tau_\phi} \frac{e^{1/\xi}}{\xi} \right), \quad \xi = \frac{\hbar}{4\pi M D_{cl}}.
\end{equation}
At not too small $\xi$ and not too large $\tau_\phi/\tau$ Eq.~\eqref{D:self} gives $D = D_{cl} - \delta D$ in agreement with Eqs.~\eqref{dD:phonons} and \eqref{D:ac}. With further increase of the absolute value of the interference correction (increase in $\tau_\phi$ or decrease in $D_{cl}$), Eq.~\eqref{D:self} shows the strong suppression of the diffusion coefficient $D\sim D_{cl}\tau/\tau_\phi \exp{(4\pi M D_{cl}/\hbar)}$. Strictly speaking, at some point the strong localization of excitons and phonon-assisted hopping should occur. However, for the experimentally relevant parameters\cite{PhysRevLett.127.076801} Eq.~\eqref{dD:phonons} should hold with minor deviations. On the other hand, another possible scenario is the polaron formation that may drastically change excitonic transport. These effects require a separate study.

In conclusion, we have developed the theory of exciton weak localization effect in two-dimensional semiconductors for the conditions where excitons mainly interact with long-wavelength acoustic phonons. We have considered interaction with longitudinal acoustic phonons with linear dispersion and with flexural phonons with quadratic dispersion. We have calculated the exciton scattering and dephasing rates and the resulting diffusion coefficients. We have identified the regimes where the semiclassical diffusion coefficient of excitons is temperature independent that makes weak localization effect particularly pronounced. The exciton-free carrier scattering effect on the weak localization has also been analyzed. The phenomena beyond the first-order quantum corrections to the exciton diffusivity have been briefly discussed.

\begin{acknowledgments}
The authors thank A. Chernikov for valuable discussions. Financial support by RSF project No.~19-12-00051 is  acknowledged (theory of exciton-phonon interactions). Z.A.~Iakovlev is also grateful to the Foundation for the Advancement of Theoretical Physics and Mathematics ``BASIS''. S.~Refaely-Abramson acknowledges support from the Israel Science Foundation, Grant No. 1208/19.
\end{acknowledgments}


%

\end{document}